\documentclass[letterpaper, 10 pt, conference]{ieeeconf}  

\IEEEoverridecommandlockouts                              
\usepackage{color}
\usepackage{amssymb}  

\usepackage{amsmath,amsfonts,amsthm}
\usepackage{mathtools}
\usepackage{booktabs}
\usepackage{tabularx}
\usepackage{arydshln}
\newlength\subtabdist
\usepackage[utf8]{inputenc}
\usepackage{graphicx}
\usepackage{xcolor}
\usepackage{comment}
\usepackage{MnSymbol}
\usepackage[linesnumbered,ruled]{algorithm2e}
\SetAlCapNameFnt{\small}
\SetAlCapFnt{\small}
\usepackage{tikz}
\usepackage{scalefnt}
\usepackage[utf8]{inputenc}
\usepackage{diagbox}
\usepackage{subfig}

\usepackage{pgfplots} 
\pgfplotsset{compat=newest} 
\pgfplotsset{plot coordinates/math parser=false} 
\newlength\figureheight 
\newlength\figurewidth 
\usepackage[strict=true,style=english]{csquotes}
\let\labelindent\relax
\usepackage{enumitem}

\newtheorem{Problem}{Problem}

\theoremstyle{definition}

\newtheorem{Example}{Example}

\newtheorem*{Remark}{Remark}

\newtheorem{Case Study}{Case Study}
\usepackage{algorithmic}
\usepackage{epsfig}
\usepackage{graphics}
\usepackage{comment}
\makeatletter
\newenvironment{subproblem}[1]{%
  \def\subtheoremcounter{#1}%
  \refstepcounter{#1}%
  \protected@edef\theparentnumber{\csname the#1\endcsname}%
  \setcounter{parentnumber}{\value{#1}}%
  \setcounter{#1}{0}%
  \expandafter\def\csname the#1\endcsname{\theparentnumber.\Alph{#1}}%
  \ignorespaces
}{%
  \setcounter{\subtheoremcounter}{\value{parentnumber}}%
  \ignorespacesafterend
}
\makeatother
\newcounter{parentnumber}

\title{\LARGE \bf{
Recurrent Neural Network Controllers for Signal Temporal Logic Specifications Subject to Safety Constraints
}}
\author{Wenliang Liu, Noushin Mehdipour and Calin Belta
\thanks{The authors are with Boston University, Boston, MA, USA (wliu97@bu.edu, noushinm@bu.edu, cbelta@bu.edu). This work was partially supported by the NSF under grant IIS-1723995}%
}

\begin{document}

\maketitle
\thispagestyle{empty}
\pagestyle{empty}

\begin{abstract}
\label{sec:abstract}
We propose a framework based on Recurrent Neural Networks (RNNs) to determine an optimal control strategy for a discrete-time system that is required to satisfy specifications given as Signal Temporal Logic (STL) formulae. RNNs can store information of a system over time, thus, enable us to determine satisfaction of the dynamic temporal requirements specified in STL formulae. Given a STL formula, a dataset of satisfying system executions and corresponding control policies, we can use RNNs to predict a control policy at each time based on the current and previous states of system. We use Control Barrier Functions (CBFs) to guarantee the safety of the predicted control policy. We validate our theoretical formulation and demonstrate its performance in an optimal control problem subject to partially unknown safety constraints through simulations.
\end{abstract}

\begin{keywords}
Optimal control; Neural networks; Autonomous systems
\end{keywords}
\section{Introduction}
\label{sec:intro}
Due to their expressivity and similarity to natural languages, temporal logics \cite{baierBook} have been used to formalize specifications for cyber-physical systems. Control policies enforcing the satisfaction of such specifications have been derived~\cite{tabuada2009verification,belta2017formal}. Our focus in this paper is Signal Temporal Logic (STL) \cite{maler2004monitoring}, which is interpreted over real-valued signals. 
STL is equipped with quantitative semantics, known as robustness, that measures how strongly a signal satisfies a specification \cite{donze2010robust}. This allows to map the problem of controlling a system under a STL specification to an optimization problem with robustness as cost function \cite{raman2014model,belta2019formal}. 
Optimizing the robustness, whether through a Mixed Integer Programming (MIP) encoding \cite{raman2014model} or a gradient-based method \cite{husam,Haghighi2019,mehdipour2019arithmetic,varnai2020robustness,gilpin2020smooth}, can be computationally expensive and might not meet real-time requirements in practice. Moreover, the optimization may converge to local optima, which might not satisfy the STL specification. 


To address these limitations, we propose a Recurrent Neural Network (RNN) controller design for a dynamical system with specifications given as STL formulae. The input to the RNN is the current state of the system and the output is the control that is predicted to maximize the STL robustness at that state. The RNN is trained using imitation learning \cite{argall2009survey}, in which the dataset consists of samples (system executions) generated by solving an optimization problem. A shallow RNN requires limited computations, and thus, it can be used for real-time control. Moreover, convergence can be improved by excluding samples with robustness scores less than a specified threshold from the dataset.

Employing neural networks (NN) in temporal logic control was proposed recently. In \cite{yaghoubi2019worst}, the authors used a feedforward NN as a feedback controller to study worst-case satisfaction of STL specifications. The feedforward NN predicted the controller at each time only based on the current state of the system. However, in general, the satisfaction of a STL specification is \emph{history-dependent}. 
For example, if a specification requires an agent to visit region A and then region B, it is not possible for the agent to know whether it should move towards B given only the current position - it needs to know whether it has visited $A$ already. For Linear Temporal Logic (LTL), the history-dependence is addressed by translating the formulae into automata 
that contain history information \cite{belta2019formal}. 
The authors of \cite{li2019formal} translated (truncated) LTL specifications into a 
finite-state automata and used reinforcement learning to train a feedforward NN for predicting satisfying control policies. 
However, STL is not equipped with such an automaton. \cite{aksaray2016q} proposed a fragment of STL such that the progress towards satisfaction could be checked with a partial trajectory, and used Markov Decision Processes (MDP) and  
Q-learning to infer control policies. Besides the restriction on the STL structure, this work also required the initial partial trajectory to be known. Most recently, \cite{leungback} used a RNN-like recurrent computation graph to compute robustness of STL formulae. By allowing back-propagation of robustness gradients, a controller was synthesized to satisfy a STL formula. 

RNNs have internal states (memory) units that can store history. In this paper, we propose a feedback RNN controller, which predicts the control policy at each state based on the current state and the history of the system, to address the history-dependence of STL satisfaction.
One important advantage of a feedback controller is its tolerance to disturbance. We demonstrate that the feedback structure of RNNs allows us to handle system disturbance and safety requirements that were not known previously (during training). These are enforced using Control Barrier Functions (CBF) \cite{ames2019control}. 
This idea is related to \cite{li2019formal}, where CBFs were used as shields to guarantee safety for both training and execution phases of a reinforcement learning framework. The authors of \cite{yaghoubi2020training} also trained a NN-based controller using imitation learning with CBF safety requirements. In contrast to our work, which uses RNN to accomplish STL specifications, \cite{yaghoubi2020training} did not consider temporal logic specifications, and the NN was solely used to solve an optimization problem with CBF constraints in a reachability problem. 

\section{Notation and Preliminaries}
\label{sec:preliminaries}

\subsection{Signal Temporal Logic (STL)}
\label{sec:prelim_STL}

An $n$-dimensional real-valued signal is denoted as $S=s_0s_1\ldots\ $, where $s_k\in\mathbb R^n$, $k\in\mathbb Z_{\geq0}$.  
The STL {\em syntax} \cite{maler2004monitoring} is defined and interpreted over $S$:\vspace{-4pt}
\begin{equation}
\label{eq:syntax}
\varphi:=\top|\;\mu \; | \; \neg\varphi \; | \; \varphi_1\land\varphi_2 \; |   \; \mathbf{F}_{I} \varphi\;| \; \mathbf{G}_{I} \varphi,
\vspace{-2pt}
\end{equation}
where $\varphi$, $\varphi_1$, $\varphi_2$ are STL formulae, $\top$ is the logical \textit{True}, $\mu$ is a \textit{predicate} over signals, $\lnot$ and $\land$ are the Boolean \textit{negation} and  \textit{conjunction} operators. The Boolean constant $\bot$ (\textit{False}) and \textit{disjunction} $\lor$ can be defined from $\top$, $\lnot$, and $\land$ in the usual way. $\mathbf{F}$ and $\mathbf{G}$ are temporal \textit{eventually} and \textit{always} operators. 
$I=[a,b]=\{k\in\mathbb Z_{\geq0}\ |\ a\leq k\leq b;\ a,b\in \mathbb Z_{\geq0}\}$ denotes a bounded time interval. $\mathbf{F}_{I}\varphi$ is satisfied if \enquote{$\varphi$ becomes \textit{True} at some time in $I$} while $\mathbf{G}_{I}\varphi$ is satisfied if \enquote{$\varphi$ is \textit{True} at all times in $I$}. 
Predicates are of the form $\mu:=l(s_k)\geq 0$, where $l: \mathbb{R}^n \to \mathbb{R}$ is a Lipschitz continuous function. 

The STL {\em qualitative semantics} determines \textit{whether} a signal $S$ satisfies a given specification $\varphi$, i.e., $S\models\varphi$, or not, i.e., $S\not\models\varphi$. Its {\em quantitative semantics}, or \textit{robustness}, assigns a real value to measure \textit{how much} a signal satisfies or violates $\varphi$. Multiple functionals have been proposed to capture the STL quantitative robustness \cite{donze2010robust,Haghighi2019,varnai2020robustness,gilpin2020smooth}. In this paper, we use the Arithmetic-Geometric Mean (AGM) robustness \cite{mehdipour2019arithmetic} which is a \emph{sound} score, i.e., a strict positive robustness indicates satisfaction of the specification, and a strict negative robustness indicates violation. However, the frameworks presented in this paper are applicable to all robustness functionals in literature. As opposed to the traditional robustness \cite{donze2010robust}, which only captures the most extreme satisfaction (or violation), AGM employs arithmetic and geometric means over all the satisfying (or violating) sub-formulae and time points in a formula and can highlight the level and frequency of satisfaction. We denote the AGM robustness of $\varphi$ at time $k$ with respect to signal $S$ by $\eta(\varphi,S,k)$. For brevity, we denote $\eta(\varphi,S,0)$ by $\eta(\varphi,S)$. The time horizon of a STL formula $\varphi$ denoted by $hrz(\varphi)$ is the smallest time point in the future for which signal values are needed to compute the robustness at the current time \cite{dokhanchi2014line}. 
\begin{figure*}[!thb]
  \centering
  \includegraphics[scale=0.40]{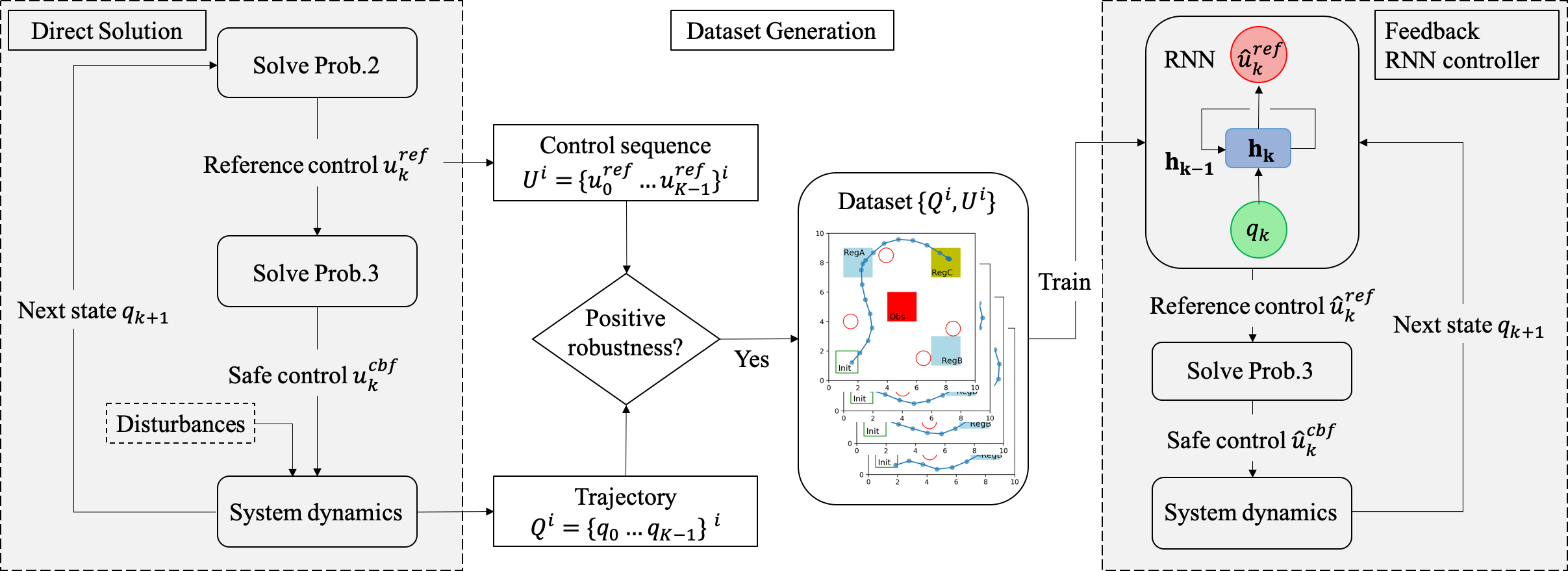}
  \caption {\small Overall approach: {\bf Left:} Safe trajectories are generated using gradient-based optimization and CBF; {\bf Middle:} Safe and satisfying trajectories (with positive robustness) and the corresponding reference controls are added to a state-control dataset; {\bf Right:} a RNN is trained on the dataset to predict reference controls for STL satisfaction. A safe feedback RNN controller is synthesized using CBF.}
  \label{fig:overview}
  \vspace{-8pt}
\end{figure*}

\subsection{Discrete-time Dynamics and Control Barrier Functions} \label{sub sec:CBF}

Consider a discrete-time control system given by\vspace{-2pt}
\begin{equation}
 \label{eq:dynamics}
  q_{k+1}=f(q_k, u_k),
  \vspace{-2pt}
\end{equation}
where $q_k \in \mathcal{Q}\subset \mathbb R^n$ is the state 
($q_0$ is the initial state) and $u_k\in \mathcal {U} \subset \mathbb R^m$ is the control input at time $k$, 
and $f:\mathcal{Q}\times\mathcal U\rightarrow\mathcal{Q}$ is a Lipschitz continuous function. Let $u_{0:K-1}$ denote the control sequence $u_{0}\ldots u_{K-1}$. The system trajectory $q_{0}q_{1}\ldots q_{K}$ generated by applying $u_{0:K-1}$ starting at $q_0$ is denoted by $\boldsymbol q(q_{0},u_{0:K-1})$.

Let $b:\mathbb R^n \rightarrow \mathbb R$. The set $\boldsymbol C=\{q\in\mathbb R^n\ |\ b(q)\geq0\}$ 
is called (forward) {\em invariant} for system 
\eqref{eq:dynamics} if all its trajectories remain in $\boldsymbol C$ for all times, if they originate in $\boldsymbol C$. 

The function $b$ is a (discrete-time, exponential) {\em Control Barrier Function} (CBF) \cite{li2019formal,agrawal2017discrete} for system \eqref{eq:dynamics} if there exist $\alpha\in[0,1]$ and $u_k\in\mathcal U$ such that:
\begin{equation}\label{eq:cbf2}
\begin{array}{c}
    b(q_0)\geq0\\
    b(q_{k+1}) + (\alpha-1)b(q_k)\geq0,\quad \forall k\in\mathbb Z_{\geq0},
    \end{array}
\end{equation}
where $q_{k+1}$, $q_k$, and $u_k$ are related by \eqref{eq:dynamics}.
The set $\boldsymbol C$ is invariant for system \eqref{eq:dynamics} if there exists a CBF $b$ as \eqref{cbf2}. This invariance property is usually referred to as {\em safety}. In other words, the system is safe if it stays inside the set $\boldsymbol C$.

\section{Problem Statement and Approach}
\label{sec:problem}
Consider system \eqref{eq:dynamics} starting at $q_0 \in \mathbb R^n$ and a differentiable cost function $J(u_k,q_{k+1})$ representing the cost of ending up at state $q_{k+1}$ by applying control input $u_k$ at time $k$.
Assume that temporal logic requirements are given by a STL formula $\varphi$ interpreted over the system states $q_0\ldots q_K$ where $K$ is the final planning horizon. For simplicity, we assume that $K=hrz(\varphi)$. However, $K$ could be any integer greater than or equal to $hrz(\varphi)$. Suppose there are $N$ safety requirements given as CBF constraints $b_i(q_k)>0$ (see Sec.\ref{sub sec:CBF}), where $i=1,\ldots,N$, $k=0,\ldots,K$.  
Let $\mathbf{b}:\mathbb R^n\rightarrow \mathbb R^N$, where $\mathbf{b}=(b_1,\ldots,b_N)$, and $\mathbf{b}(q_k)>0$ is interpreted componentwise. Our goal is to find a control policy for system \eqref{eq:dynamics} that maximizes satisfaction of the STL specification, minimizes the cost function and satisfies the safety requirements. 
\vspace{-3pt}
\begin{Problem} 
\label{pb:whole}
Given system dynamics \eqref{eq:dynamics}, cost function $J$, STL formula $\varphi$, initial state $q_0$ and safety requirement $\mathbf{b}(q_k)>0$, find an optimal control policy $u_{0:K-1}^{*}$ that maximizes robustness and minimizes the penalized cost:\vspace{-4pt}
\begin{equation}
\label{eq:whole}
\begin{aligned}
u_{0:K-1}^{*}=\arg\max_{u_{0:K-1}} \; & \eta(\varphi, \boldsymbol q(q_{0},u_{0:K-1})) -\lambda\sum_{k=0}^{K-1}J(u_k,q_{k+1})\\
\textrm{s.t.} \quad &u_k\in \mathcal U \subset \mathbb R^m,\ k=0,\ldots,K-1\\
  &q_{k+1}=f(q_k,u_k),\ k=0,\ldots,K-1\\
  &\mathbf{b}(q_{k})>0 ,\ k=0,\ldots,K\\
\end{aligned}
\vspace{-4pt}
\end{equation}
where $\lambda$ captures the trade-off between satisfying the specification $\varphi$ and minimizing the cost.
\end{Problem}

The solution to Pb.~\ref{pb:whole} is an open loop controller, as the synthesized control sequence is applied to the entire planning horizon. This formulation would fail to satisfy the specifications if the actual system trajectory deviates from the synthesized one due to the existence of disturbances in the system dynamics or changes in the safety constraints (e.g., moving obstacles). Instead, we propose to solve Pb.~\ref{pb:whole} by finding the optimal control at each time based on the current and past\footnote{state history is necessary to decide STL satisfaction, see Secs. \ref{sec:intro} and \ref{sec:prelim_STL}}
states of the system, which gives a history-dependent state feedback controller. 
Specifically, at each time $k$, the optimization variable $u_{k:K-1}$ covers the rest of the time and the feedback information includes the current state $q_k$ and the \emph{history 
trajectory} $q_0\ldots q_{k-1}$ (this property is called \emph{history-dependence} of STL).
However, solving the optimization problem at each time is time-consuming, which is a problem for real-time implementations. Moreover, the optimization may converge to a local optimum (negative robustness). We address these limitations by training a RNN to predict the control policy at each time (details in Sec. \ref{sec:synthesis}). Neural networks execute very fast. They can take a long time to train, but this computation is performed off-line (before deployment). Our goal is to make the RNN controller flexible, i.e., we want the trajectories generated from the predicted RNN control input to be able to meet the STL specifications under various safety constraints (e.g., unforeseen or dynamic safety constraints), without a need to re-train the RNN when the safety constraints change. %
\section{Reference Control and Safe Control}
\label{sec:reform}

In order to generate a dataset for a flexible RNN, we decompose the optimization problem at each time into two problems: Pb. \ref{pb:ref} and Pb. \ref{pb:cbf}. The solution to Pb. \ref{pb:ref} provides a \emph{reference control} sequence that gives the ``direction" towards the satisfaction of the STL formula but does not consider the safety constraints. In Pb. \ref{pb:cbf}, the first control input (input at the current time) in the reference control sequence is modified (if needed) using CBFs to provide a \emph{safe control} which is applied to the system to move to the next state. Pb. \ref{pb:ref} and Pb. \ref{pb:cbf} are recursively solved at each time until the final time is reached, as shown in Fig. \ref{fig:overview} (left). At each time, the current (safe) system state and the (possibly unsafe) reference control are added to ordered sequences of previous states and previous reference controls, respectively. At the final time, the two sequences are combined as a data pair to generate a state-control dataset, on which the RNN is trained (middle of Fig. \ref{fig:overview}). This framework enables the RNN to predict the reference control, i.e., the solution to Pb. \ref{pb:ref}, at each time based on the current state and the history trajectory. The predicted reference control drives the next state of the system towards STL satisfaction, and is modified by solving Pb. \ref{pb:cbf} to ensure it is safe as shown in Fig. \ref{fig:overview} (right).

There are two main advantages of training the RNN on the reference control (instead of the safe control) and using CBF to guarantee safety of the RNN controller. First, we can 
accommodate safety constraints different from those in the dataset. Otherwise, if the RNN was trained on the safe control, it would assume the safety constraints in the dataset used for training always exist. Second, the final trajectory is guaranteed to be safe independent of the performance of the RNN. Even though safety of the predicted control input is guaranteed after RNN by solving Pb. \ref{pb:cbf}, we still solve Pb. \ref{pb:cbf} during dataset generation to enlarge the search space (i.e., explore more states that might appear due to various safety constraints and include more state-control data in the dataset). 

We propose two versions of Pb. \ref{pb:ref} - either can be used depending on the structure and length of the STL formula. 

\begin{figure}[t]
    \centering
  \subfloat[\label{fig:eg4}]{\includegraphics[width=2.8cm]{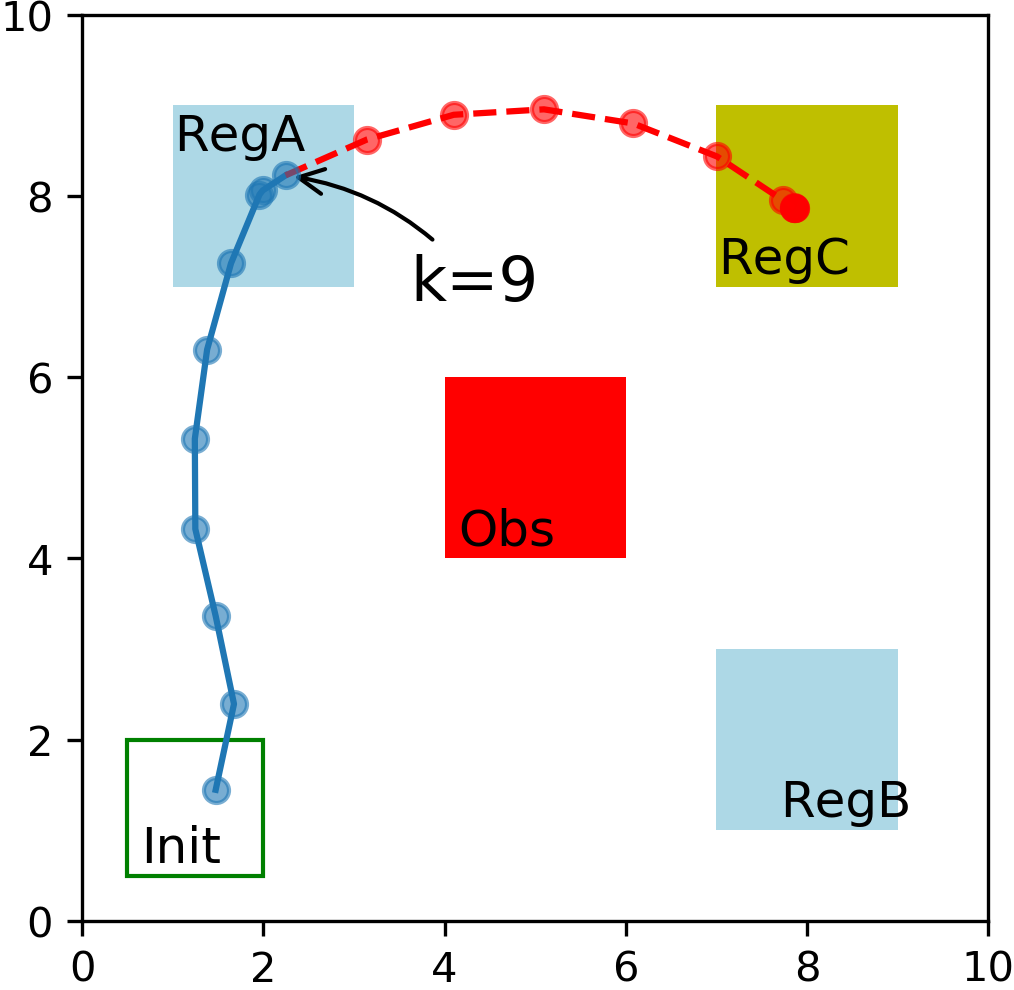}}\ 
  \subfloat[\label{fig:eg5}]{\includegraphics[width=2.8cm]{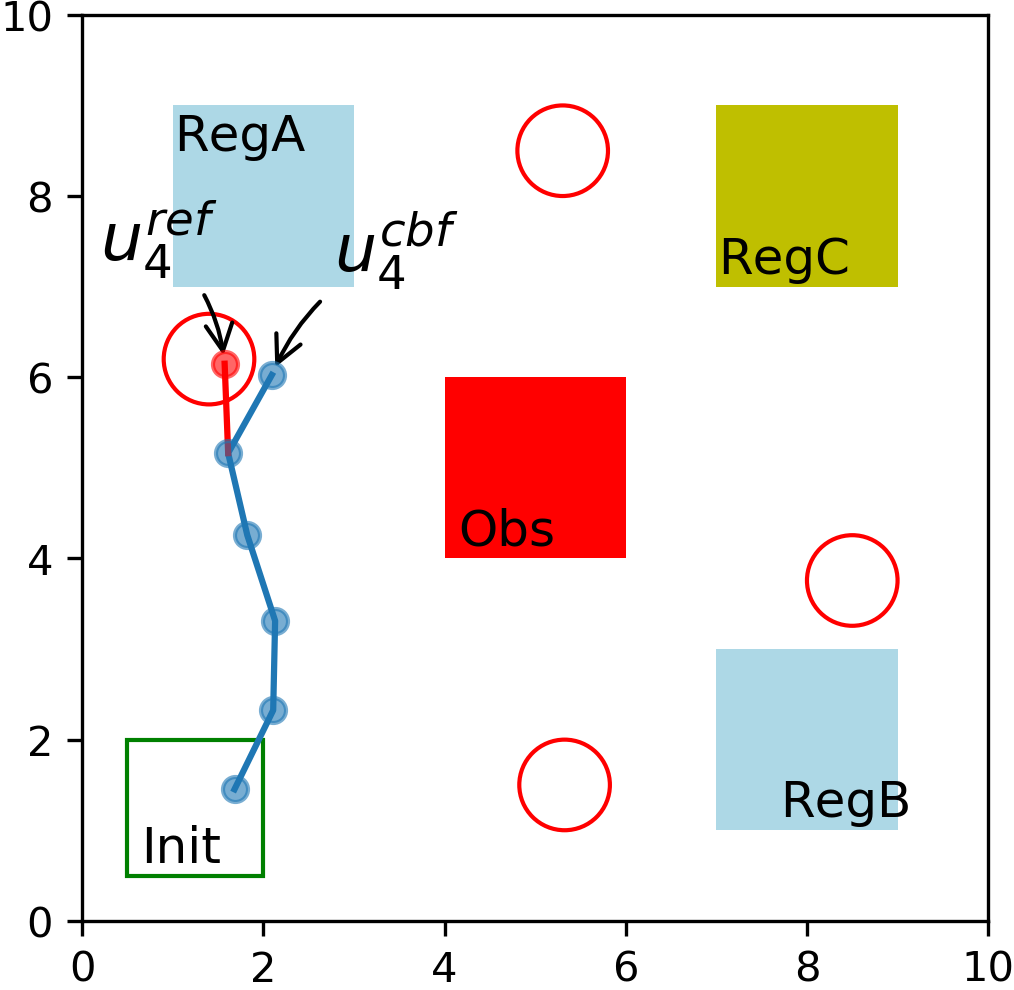}}\
  \subfloat[\label{fig:eg6}]{\includegraphics[width=2.8cm]{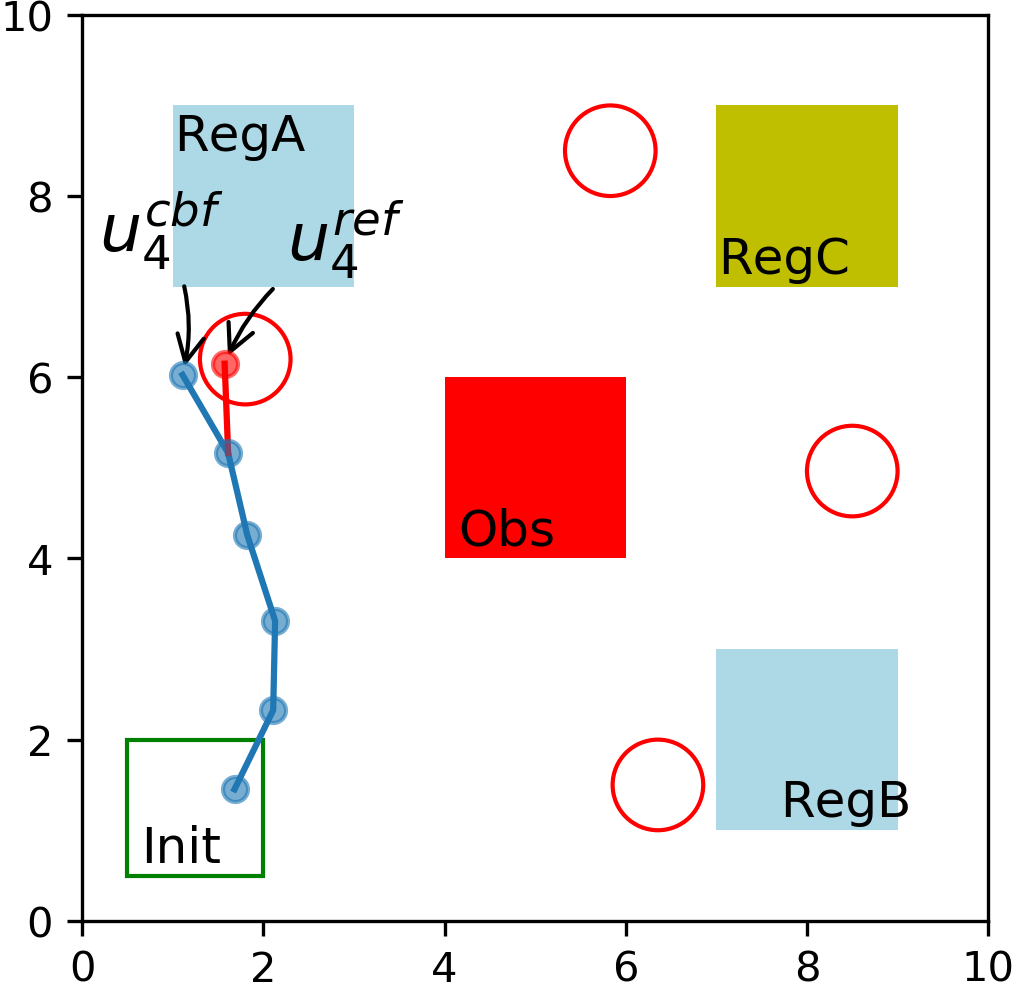}}
  \caption{\small (a): History trajectory (blue) and trajectory to be optimized (red) at time $k=9$. (b) and (c): Reference control $u_4^{ref}$ at time $k=4$ steering the robot to the red point and safe control $u_4^{cbf}$ steering the robot to the blue point. The history trajectory, current state $q_4$, and reference control $u_4^{ref}$ are the same in (b) and (c). The positions of the obstacles are different, which result in different $u_4^{cbf}$.}
  \label{fig:eg2} 
  \vspace{-8pt}
\end{figure}

\begin{subproblem}{Problem}\label{pb:ref}
\vspace{-3pt}
\begin{Problem}[Reference Control] \label{pb:refA}
Given system dynamics \eqref{eq:dynamics}, cost function $J$, STL formula $\varphi$, current state $q_k$ and history trajectory $q_0\ldots q_{k-1}$, reference control $u_{k:K-1}^\text{ref}$ at time $k\in[0,K-1]$ is found by:
\begin{equation}
\vspace{-5pt}
\label{eq:refA}
\begin{aligned}
u_{k:K-1}^\text{ref}=\arg\max_{u_{k:K-1}} \quad & \eta(\varphi, q_0\ldots q_{k-1}\boldsymbol q(q_k,u_{k:K-1})) \\&-\lambda\sum_{j=k}^{K-1}J(u_j,q_{j+1})\\
\textrm{s.t.} \quad &u_j\in \mathcal{ U} \subset \mathbb R^m, j=k,\ldots,K-1\\
  &q_{j+1}=f(q_j,u_j), j=k,\ldots,K-1\\
\end{aligned}
\end{equation}
\vspace{-4pt}
\end{Problem}
By solving Pb. \ref{pb:refA} at time $k$, we find a reference trajectory $\boldsymbol q(q_k,u_{k:K-1})$ which along with the history trajectory satisfies the STL formula, i.e., $q_0\ldots q_{k-1}\boldsymbol q(q_k,u_{k:K-1}) \models \varphi$. 

\begin{Example}
\label{eg:2}
Consider a robot in a 2-dimensional workspace in Fig. \ref{fig:eg4}. The specification is to \enquote{\emph{eventually} visit $RegA$ \emph{or} $RegB$ within  [1,10] \emph{and} \emph{eventually} visit $RegC$ within [11,20] \emph{and} \emph{always} avoid $Obs$}, 
written as a STL formula: \vspace{-2pt}
\begin{equation}
\label{stl}
\begin{aligned}
    \varphi_1=(\boldsymbol{\rm F}_{[1,10]}(RegA \lor RegB))\land(\boldsymbol{\rm F}_{[11,20]}RegC)\\
    \land(\boldsymbol{\rm G}_{[0,20]}\lnot Obs),
\end{aligned}
\vspace{-2pt}
\end{equation}
with $hrz(\varphi_1)=20$. Consider the trajectory from Fig. \ref{fig:eg4}, and (current) state $q_9$ at time $k=9$. The blue trajectory $q_0\ldots,q_8$ is the history trajectory, and the red trajectory $q_{10}\ldots q_{20}$ is the synthesized trajectory from the solution of Pb. \ref{pb:refA}. 
\end{Example}

If the horizon of $\varphi$ is large, Pb. \ref{pb:refA} may become prohibitively expensive. If $\varphi=\boldsymbol{\rm G}_{[0,k_1]}\phi$, we can use a model predictive control (MPC) approach \cite{sadraddini2015robust} to shorten the optimization (planning) horizon. Let $h^\phi = hrz(\phi)$ and let $h_p$ denote the (shorter) prediction horizon. Instead of optimizing the entire trajectory over $K=k_1 + h^\phi$ steps, in a MPC framework, we optimize the trajectory for the next $H=h_p +h^\phi$ steps by recursively maximizing the robustness of $\boldsymbol{\rm G}_{[0,h_p ] }\phi$ with respect to the partial trajectory $\boldsymbol q(q_k,u_{k:k+H-1})$, $k=0,1,\ldots,K-H$. For example, at time $k=0$, we maximize the robustness of $\boldsymbol{\rm G}_{[0,h_p ]}\phi$ with respect to $q_0,q_1,\ldots,q_{H}$; at $k=1$, we maximize the robustness of $\boldsymbol{\rm G}_{[0,h_p ]}\phi$ with respect to  $q_1,q_2,\ldots,q_{H+1}$, etc. We need to ensure that, when moving forward, the satisfaction of $\phi$ that was obtained during the previous optimizations still holds. Therefore, when maximizing the robustness of $\boldsymbol{\rm G}_{[0,h_p ]}\phi$ with respect to the partial trajectory starting from time $k$, we need to enforce the robustness of $\phi$ to remain positive at the previous $h^{\phi}-1$ steps \cite{sadraddini2015robust}. Formally, we have:
\begin{Problem}[Reference Control using MPC] \label{pb:refB}
\vspace{-3pt}
At time $k\in[h^\phi-1,K-H]$, given system dynamics \eqref{eq:dynamics}, cost function $J$, STL formula $\varphi=\boldsymbol{\rm G}_{[0,k_1]}\phi$, current state $q_k$ and history trajectory $q_{k-h^\phi+1} \ldots q_{k-1}$, reference control $u_{k:k+H-1}^\text{ref}$ is found by\footnote{Note that, when $k<h^\phi-1$ or $k>K-H$, the corresponding horizons in \eqref{eq:refB} need to be modified \cite{sadraddini2015robust}.}:\vspace{-3pt}
\begin{equation}
\label{eq:refB}
\begin{aligned}
u_{k:k+H-1}^\text{ref}=&\arg\max_{u_{k:k+H-1}}  \eta(\boldsymbol{\rm G}_{[0,h_p]}\phi, \boldsymbol q(q_k,u_{k:k+H-1})) \\&-\lambda\sum_{j=k}^{k+H-1}J(u_j,q_{j+1})\\
\textrm{s.t.} \  &u_j\in \mathcal{ U} \subset \mathbb R^m,\ j=k,\ldots,k+H-1\\
  &q_{j+1}=f(q_j,u_j),\ j=k,\ldots,k+H-1\\
  &\eta(\phi,q_{k-h^{\phi}+1+i},\ldots,q_{k-1}\boldsymbol q(q_k,u_{k:k+i}))>0,\\
  &\qquad i=0,\ldots,h^{\phi}-2.
\end{aligned}
\vspace{-3pt}
\end{equation}
  \vspace{-3pt}
\end{Problem}
\end{subproblem}

The solution to Pb. \ref{pb:refA} or Pb. \ref{pb:refB} is the \emph{reference control} without considering safety constraints. The reference control at the current time $u_k^\text{ref}$ will be added to the sequence of reference controls for dataset generation, and subsequently modified to satisfy the safety constraints:
\begin{Problem}[Safe Control] \label{pb:cbf}
\vspace{-3pt} 
At time $k\in[0,K-1]$, given system dynamics \eqref{eq:dynamics}, current state $q_k$, safety constraints $\mathbf{b}(q_k)>0$, and reference control $u_{k}^\text{ref}$ (possibly unsafe), safe control policy $u_k^\text{cbf}$ is found by: 
\begin{equation}
\label{eq:cbf3}
\begin{aligned}
u_k^\text{cbf}=\arg\min_{u_k} \quad & \|{u_k-u_k^\text{ref}}\|^2 \\
\textrm{s.t.} \quad & \mathbf{b}(f(q_k,u_k))+(\alpha-1)\mathbf{b}(q_k)>0,\\
  &u_k\in \mathcal{ U} \subset \mathbb R^m\\
\end{aligned}
\end{equation}
\vspace{-8pt}
\end{Problem}
\begin{Example}
\label{eq:3}
At time $k=4$, the reference control $u_4^{ref}$, which steers the robot from Ex. \ref{eg:2} to satisfy $\varphi_1$ (go to $RegA$), is computed from Pb. \ref{pb:ref}. Assume that there are $4$ circular obstacles appearing at time $k=4$, as shown in Fig. \ref{fig:eg5} and Fig. \ref{fig:eg6}, under the reference control $u_4^{ref}$, the robot will collide with one of the obstacles. However, by solving Pb. \ref{pb:cbf}, we can modify the reference control to $u_4^{cbf}$ to avoid collision. With the same STL formula and current state and history trajectory, the reference control $u_4^{ref}$ is
determined, while the safe control $u_4^{cbf}$ depends on the different positions of obstacles (Fig. \ref{fig:eg5} and \ref{fig:eg6}). Since the positions of obstacles when testing (deploying) the RNN are unforeseen, 
we save the current state $q_4$ and the reference control $u_4^{ref}$ into the dataset to teach the RNN the reference control towards STL satisfaction. When testing the RNN, we modify its output depending on the positions of obstacles at that moment. 
\end{Example}

{\bf \noindent Direct solution} The method used to generate the dataset, which we refer to as the \emph{direct solution}, is summarized below.  At each time $k$, we solve Pb. \ref{pb:refA} or Pb. \ref{pb:refB}, depending on the structure of $\varphi$, to get a reference control sequence $u_{k:K-1}^\text{ref}$ or $u_{k:k+H-1}^\text{ref}$. We take $u_k^\text{ref}$ and modify it, if needed, by solving Pb. \ref{pb:cbf} to get the safe control input $u_k^\text{cbf}$. By applying $u_k^\text{cbf}$ to the system dynamics (also adding a disturbance $w\in\mathcal W\subset\mathbb R^n$ such that $q_{k+1}=f(q_k,u_k^{cbf})+w$ to further enlarge the exploration space), we will find the next state $q_{k+1}$, and Pb. \ref{pb:ref} and Pb. \ref{pb:cbf} are recursively solved for time $k+1$ until the final time is reached. Both Pb. \ref{pb:ref} and Pb. \ref{pb:cbf} are solved using gradient based optimization methods. 
\section{RNN Controller Synthesis}
\label{sec:synthesis}

{\bf\noindent Dataset Generation} 
Given an initial state $q_0$ and the safety constraints $\mathbf b(q_k)>0$, $k=0,\ldots, K$, we can use the direct solution to generate a safe trajectory denoted by $Q=q_0\ldots q_{K}$, and the corresponding reference control sequence denoted by $U=u^{ref}_0\ldots u^{ref}_{K-1}$. Together, $(Q,U)$ is considered as a paired state-control data. 
In order to create a dataset for RNN, we generate a set of $M$ random initial states $q_0^i$, $i=1,\ldots,M$ 
and corresponding safety constraints $\mathbf{b}^i, \; i=1,\ldots,M$. 
For each $q_0^i$ and associated $\mathbf{b}^i$, a safe trajectory $Q^i$ and corresponding reference control $U^i$ are generated. If $Q^i$ has positive robustness, i.e., $\eta(\varphi, Q^i)>0$, the state-control pair $(Q^i,U^i)$ is added to the dataset $\mathbf{D}$ (as illustrated in Fig. \ref{fig:overview}). 

{\bf\noindent Feedback RNN Controller}
Due to the \emph {history-dependence} of STL, the control at each time depends on the current state and the history trajectory. Formally, at each time $k$, $u_k^{ref}=g(q_0,\ldots,q_k)$.
Since neural networks are known to be universal function approximators, the feedback function $g$ can be approximated by a RNN with weights $(\boldsymbol {W}_1, \boldsymbol {W}_2)$: \vspace{-3pt}
\begin{equation}
    \label{RNN}
    \begin{aligned}
    \mathbf h_k &=\mathcal{R}(q_k,\mathbf h_{k-1},\boldsymbol W_1)\\
    \hat u_k^{ref} &= \mathcal{N}(\mathbf h_k,\boldsymbol W_2),
    \end{aligned}
    \vspace{-2pt}
\end{equation}
where $\mathbf h_k$ is the RNN hidden state at time $k$, which encodes the history trajectory, and $\hat u_k^{ref}$ is the RNN output, which is the predicted control policy. By passing the history trajectory as the hidden state (with variable lengths depending on the current time $k$), RNN can manage the history-dependence of the STL satisfaction.

The RNN formulated in \eqref{RNN} is trained on the state-control dataset $\mathbf D$ such that the prediction error between the reference control $u^{ref}_k$ (from the dataset) and the predicted control $\hat u^{ref}_k$ at all times $k=0,1,\ldots K-1$ is minimized:
\begin{equation}
    \min_{\boldsymbol W_1,\boldsymbol W_2} \sum_{\mathbf D}\sum_{k=0}^{K-1}\|{\mathcal{N}(\mathcal{R}(q_k,\mathbf h_{k-1},\boldsymbol W_1),\boldsymbol W_2) - u^{ref}_k}\|^2.
\end{equation}
To implement the RNN, we use a Long Short Term Memory (LSTM) network \cite{hochreiter1997long}. Similar to \cite{yaghoubi2019worst}, we also apply a hyperbolic tangent function on the RNN outputs (i.e., the predicted control inputs at each time) in order to meet the control constraints $u_k\in \mathcal U$. 

To guarantee the safety of the trajectory, Pb. \ref{pb:cbf} is solved to adjust $\hat u^{ref}_k$ and obtain a safe control $\hat u^{cbf}_k$. This safe control $\hat u^{cbf}_k$ is applied to the system to steer it to the next state $q_{k+1}$, and the process is repeated until reaching the final time.

\section{Case Studies}
\label{sec:case}

In this section, we show the efficacy of our proposed RNN framework and compare our results with the direct solution. All algorithms 
were implemented in Python running on a Mac with a 2.6GHz Core i7 CPU and 16GB of RAM. We used Sequential Quadratic Programming (SQP) \cite{bertsekas1997nonlinear} from the scipy.minimize package \cite{virtanen2020scipy} to solve Pb. \ref{pb:ref} and Pb. \ref{pb:cbf}. The RNN was implemented using the Pytorch package \cite{paszke2017automatic}. 

We present two case studies, which illustrate the proposed framework using Pb. \ref{pb:refA} (Case Study \ref{ex:obs}) and Pb. \ref{pb:refA} (Case Study \ref{cs:2}), respectively. For both,  
the cost function is defined as $J=\frac{1}{2}\sum_{k=0}^{K-1}\|{u}_k\|^2$. 
The RNN structure consists of a LSTM network with $2$ hidden layers and $64$ nodes in each layer. 
The dataset $\mathbf{D}$ contains state-control pairs $(Q,U)$ with random initial states in a fixed region. The trained RNN controller is tested on $1000$ random initial states (in the same fixed region) with random safety constraints. 

\begin{Case Study}
\label{ex:obs}

Consider the scenario from 
Ex. \ref{eg:2}, and assume the discrete-time dynamics of the robot is given by:\vspace{-2pt}
\begin{equation}
    \label{system2}
    \begin{aligned}
    x_{k+1} &=x_k+\frac{v_k}{\omega_k}\big (\sin{(\theta_k+\omega_k)}-\sin{\theta_k}\big),\\
    y_{k+1} &=y_k+\frac{v_k}{\omega_k}\big ( \cos{\theta_k}-\cos{(\theta_k+\omega_k)}\big ),\\
    \theta_{k+1} &=\theta_k+\omega_k.
    \end{aligned}
    \vspace{-2pt}
\end{equation}
$q=(x,y,\theta)$ is the state vector with position and orientation of the robot, and the control input $u = (v,\omega)$ contains the forward and angular speeds, where $v\in [0,1], \omega\in[-0.5,0.5]$. 

Besides the fixed obstacle specified in Eq. \eqref{stl}, we assume random circular obstacles emerge in the environment (see  Fig. \ref{fig:result}). 
These obstacles are considered as additional safety constraints that can be enforced by CBFs $b_i$ (from Eq. \eqref{eq:cbf2}):
\begin{equation}
\label{cbf}
    b_i(q) = (x-x_{o,i})^2+(y-y_{o,i})^2-r_{o,i}^2,\quad i=1,2,3,4
\end{equation}
where $(x_{o,i},y_{o,i})$ is the center of the $i^{th}$ circular obstacle and $r_{o,i}$ is its radius.

The procedure described in the direct solution (with Pb. \ref{pb:refA}) is applied to generate a dataset, 
considering $\lambda=0$ in \eqref{eq:refA} and $\alpha=0.7$ in \eqref{eq:cbf3}. The norm in \eqref{eq:cbf3} is also modified to $(v_k-v_k^{ref})^2+\gamma(\omega_k-\omega_k^{ref})^2$ where $\gamma=0.03$ in order to encourage the robot to turn instead of slowing down when approaching an obstacle. 
Generating a dataset of $500$ (satisfying) trajectories takes about $2$ hours, and training the RNN on this dataset for $300$ epochs takes about $2$ minutes. 

The success rate (obtaining safe and satisfying trajectories) for the RNN solution is $99.5\%$. Fig. \ref{fig:result} shows sample trajectories for random initial conditions and safety constraints (circular obstacles in Fig. \ref{fig:result}a and \ref{fig:result}b) obtained by applying the safe control $\hat u^{cbf}$. As illustrated, by separating the CBF from the RNN controller, safety constraints are guaranteed to be satisfied, even for previously unknown safety constraints, and independent of the performance of the RNN (Fig. \ref{fig:result}a, Fig. \ref{fig:result}b). Moreover, since the RNN is trained on the reference control inputs, the trajectory generated from the predicted control inputs avoids unnecessary re-directions when no additional safety constraints exist (Fig. \ref{fig:result}c).

\begin{figure}[t!]
    \centering
  \subfloat[]{\includegraphics[width=2.8cm]{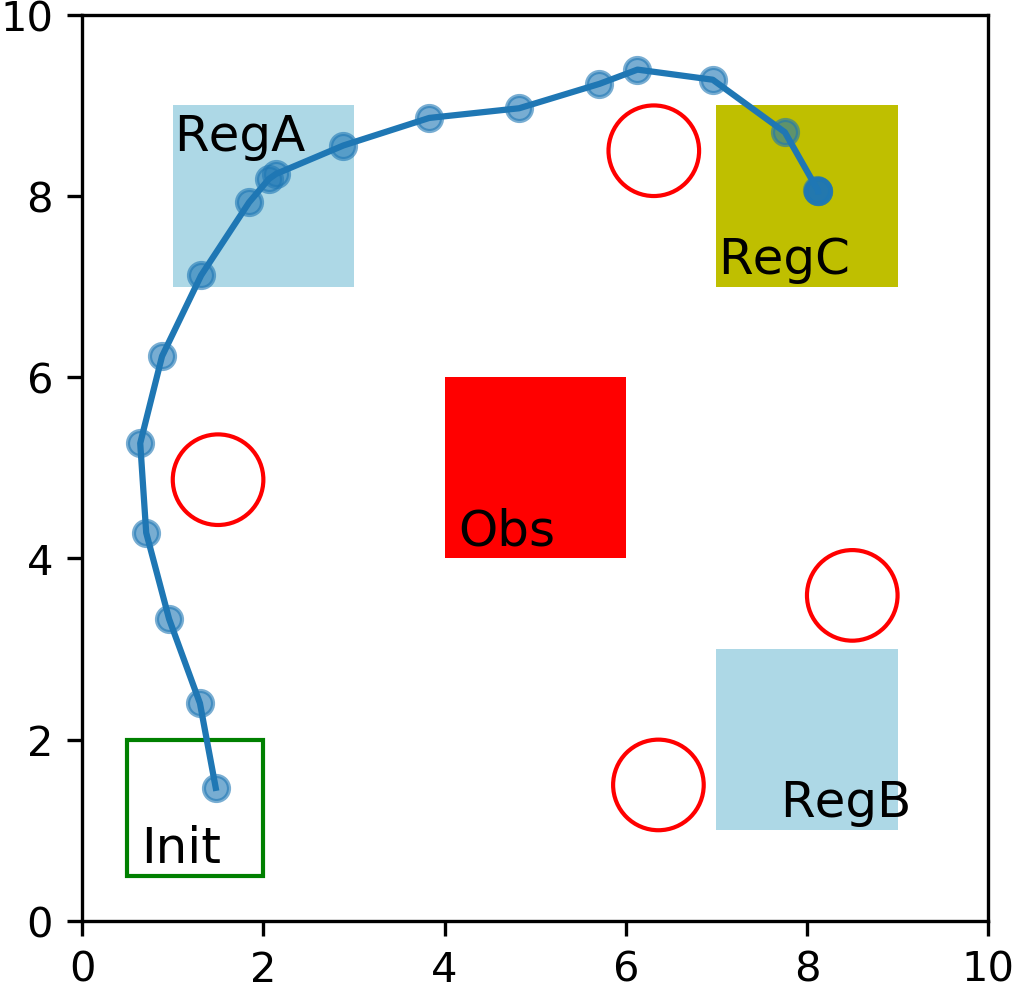}}\ 
  \subfloat[]{\includegraphics[width=2.8cm]{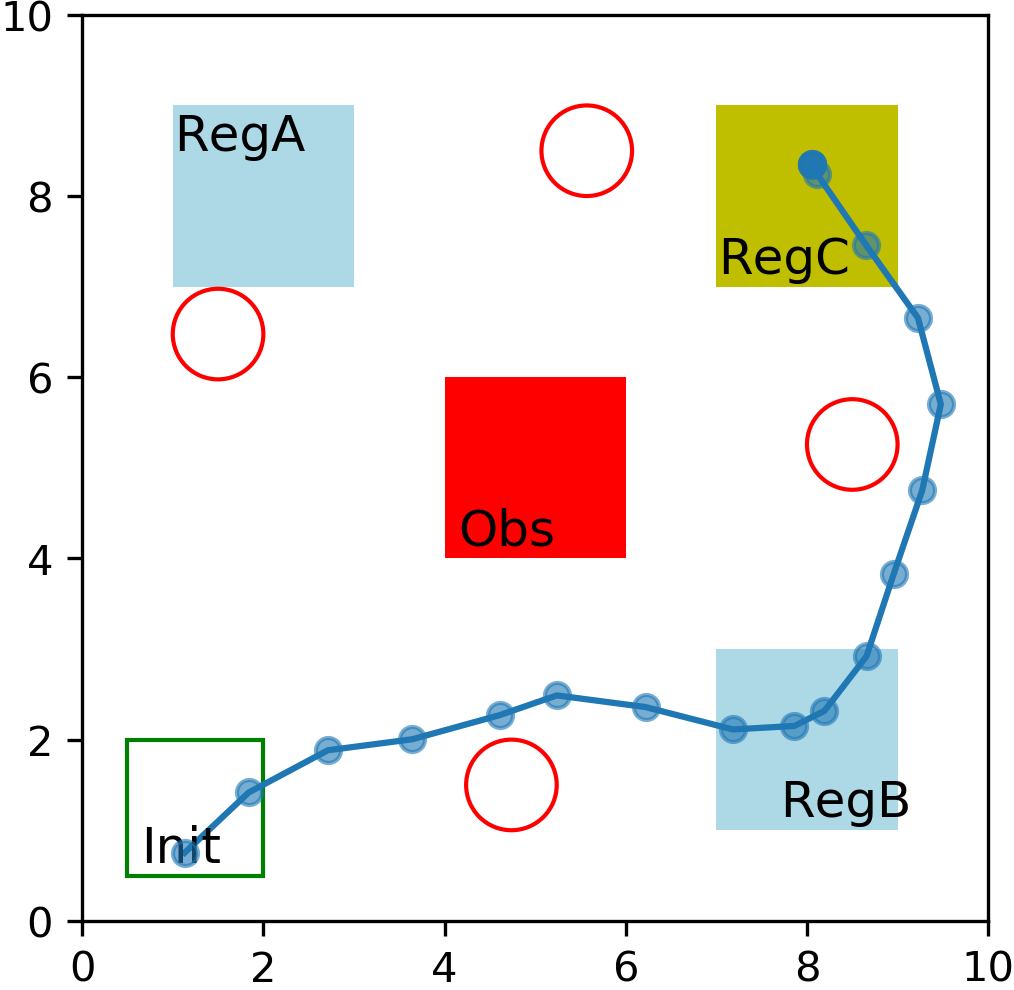}}\ 
  \subfloat[]{\includegraphics[width=2.8cm]{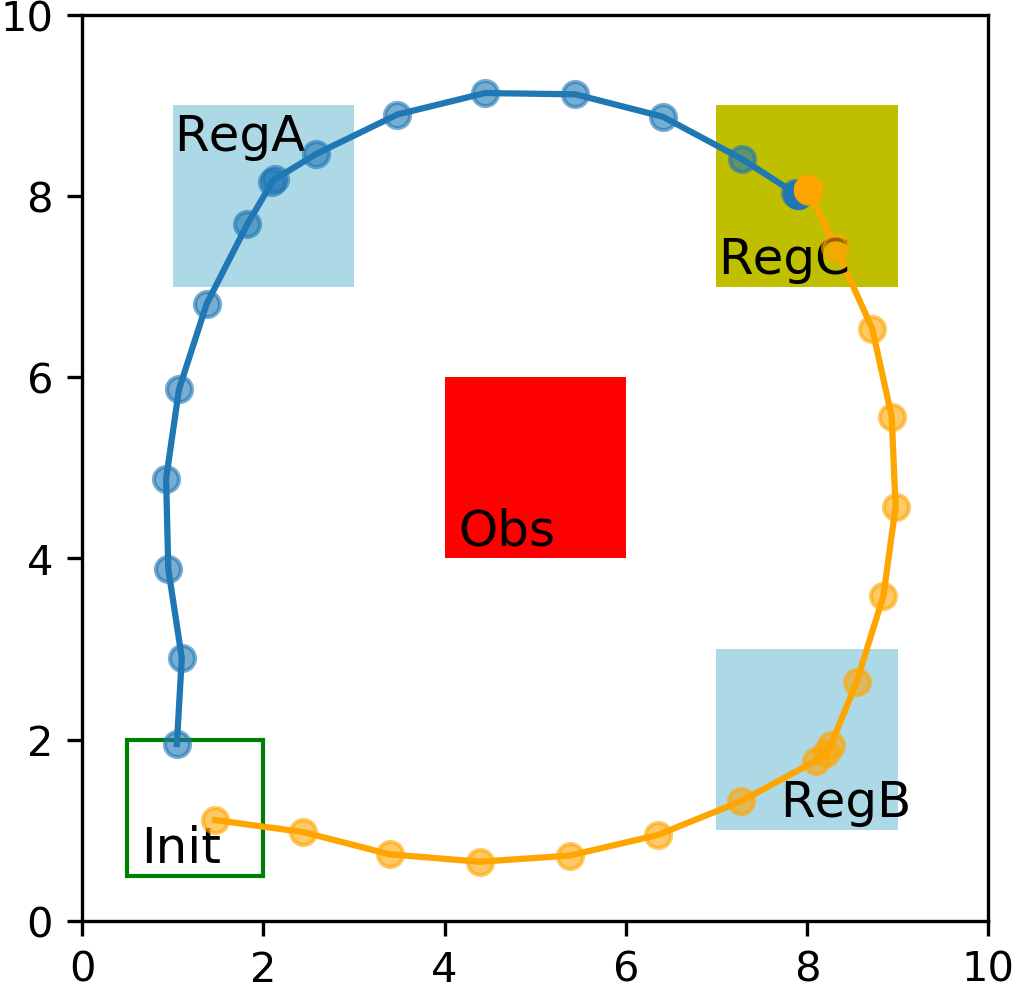}}
  \caption{\small Trajectories generated using our RNN-CBF framework. Only the solid obstacle (rectangle) was known during the RNN training. CBF guarantees safety against the random unknown obstacles (circles) if they exist.}
  \vspace{-5pt}
  \label{fig:result} 
\end{figure}

The average normalized robustness for the trajectories generated by the RNN solution for 1000 random runs is $0.0425$, and the average normalized robustness for the trajectories in dataset $\mathbf{D}$ from the direct solution (all of which are trajectories with positive robustness) is $0.0423$. Since the random obstacles serve as disturbances during dataset generation, no additional disturbances are added, hence the robustness comparison of both solutions is fair. This suggests that the performance of the RNN controller is as good as the direct solution. 
Computation times for the direct solution and the RNN solution are shown in Table \ref{tb:1}. The comparison confirms that the proposed RNN controller is 
much faster 
and suitable for real-time synthesis and planning applications. 

\begin{table}[t]
\centering
\setlength{\belowcaptionskip}{10pt}
\vspace{-10pt}
\caption{\small Computation times for the direct and RNN solutions}
\label{tb:1}
\begin{tabular}{|l|c|c|}
\hline
\diagbox{Time}{Methods}& Direct solution& RNN solution\\
\hline
Solve Pb.\ref{pb:refA} (single time)& $0.635s$& $0.000417s$\\
\hline
Generate entire trajectory& $12.8s$& $0.0582s$\\
\hline
\end{tabular}
\vspace{-6pt}
\end{table}

\end{Case Study}
\begin{Case Study}
\label{cs:2}

Consider a discrete-time system given by:\vspace{-2pt}
\begin{equation}
    \label{system3}
    \begin{aligned}
    x_{k+1} &=x_k+u_{x,k},\\
    y_{k+1} &=y_k+u_{y,k},
    \end{aligned}
    \vspace{-2pt}
\end{equation}
in a configuration shown in Fig. \ref{fig:res2}a. $q=(x,y)$ is the state vector, and $u=(u_x,u_y)$ is the control input with $\mathcal{U}=[-0.6,0.6]^2$. 
The specification is \enquote{for all times in $[0,7]$, \emph{eventually} visit $RegA$ every $3$ steps \emph{and} \emph{eventually} visit $RegB$ every $3$ steps}, which translates to the STL formula:    \vspace{-3pt}
\begin{equation}
\label{stl2}
    \varphi_2=\boldsymbol{\rm G}_{[0,7]}(\boldsymbol{\rm F}_{[0,3]}RegA \land \boldsymbol{\rm F}_{[0,3]}RegB).
        \vspace{-2.5pt}
\end{equation}
With $\phi=\boldsymbol{\rm F}_{[0,3]}RegA \land \boldsymbol{\rm F}_{[0,3]}RegB$, we have $h^{\phi}=3$. Let $q_0$ be a random position inside $RegB$. We use Pb. \ref{pb:refB} to find reference control inputs and generate a dataset $\mathbf{D}$ based on the direct solution procedure. In this example, we set $h_p=0$, $\lambda=10^{-6}$, and $\alpha=0.8$. 
We also add a random disturbance $w\in [-0.05,0.05]^2$ to the system dynamics when generating the dataset. 
Safety is specified as a circular region (Fig. \ref{fig:res2}): \vspace{-3pt}
\begin{equation}
\label{cbf2}
    b(q) = -(x-x_{safe})^2-(y-y_{safe})^2+r_{safe}^2,
    \vspace{-3pt}
\end{equation}
with $(x_{safe},y_{safe})$ and $r_{safe}$ being its center and radius.

Generating a dataset of $1000$ satisfying trajectories takes about $40$ minutes and training the RNN for $300$ epochs takes about $2$ minutes. Fig. \ref{fig:res2}a shows a sample trajectory obtained by applying the safe control $\hat u^{cbf}$. As illustrated in Fig. \ref{fig:res2}b, the system periodically visits $RegA$ and $RegB$ every $3$ steps. In this example, the RNN controller produces satisfying trajectories with a success rate of $100\%$. The computation times for the direct solution and RNN solution are $2.252s$ and $0.00885s$, respectively, which also illustrates the advantages of the RNN controller for real-time applications. 

\begin{figure}[htb]
\centering
\subfloat[]{\includegraphics[height=3cm]{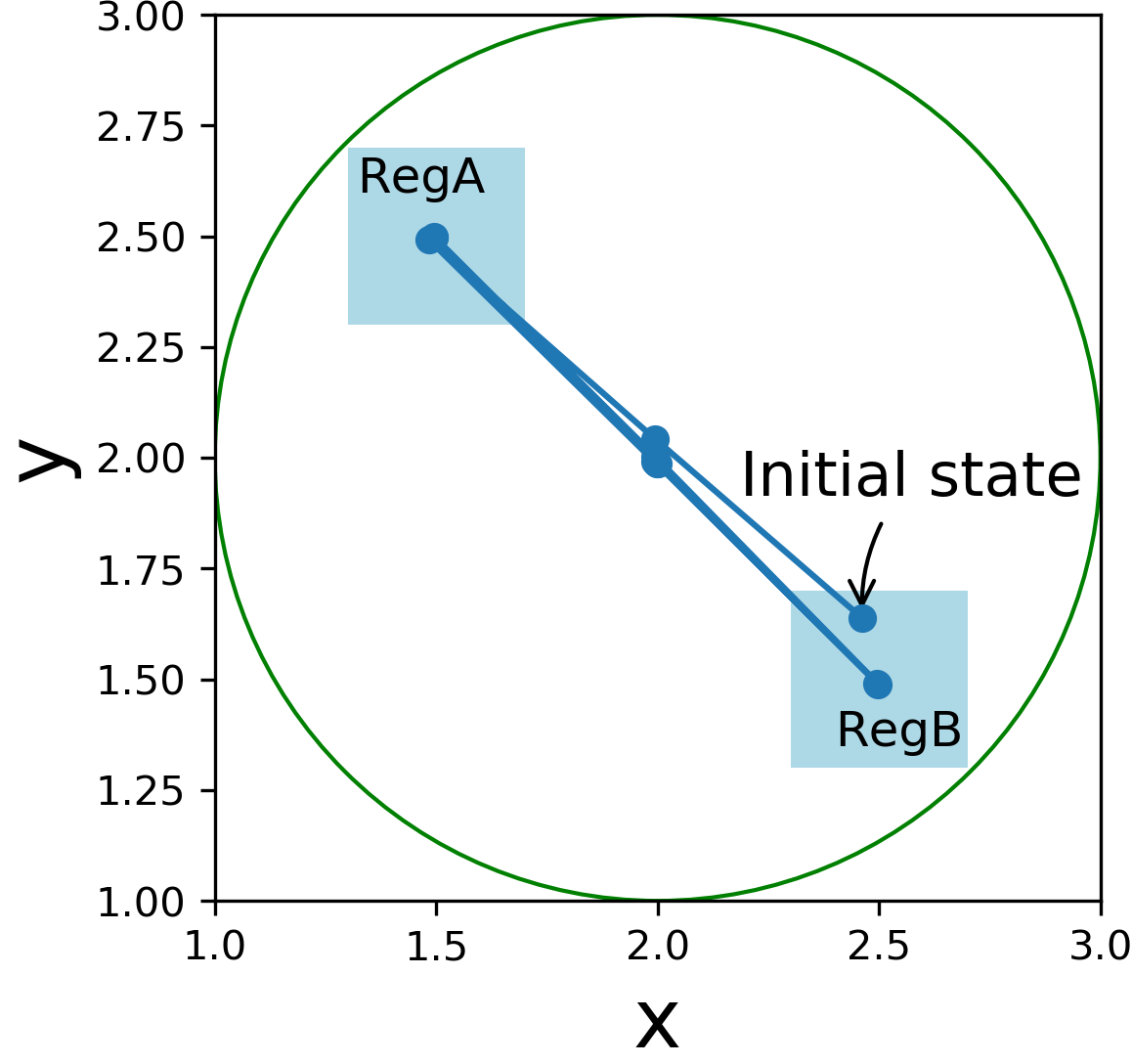}}\
\subfloat[]{\includegraphics[height=3cm]{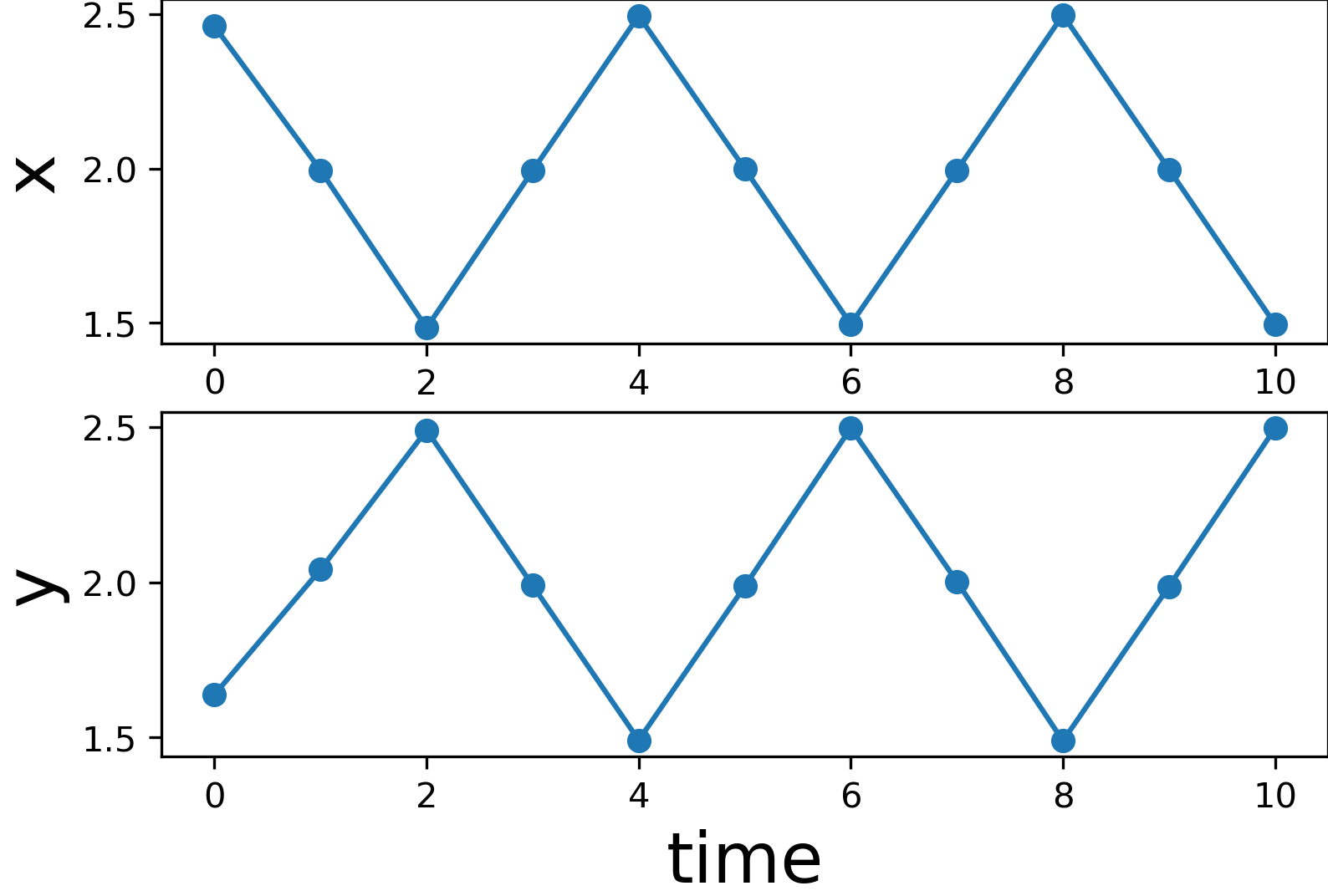}}
\caption{\small A trajectory generated using our RNN-CBF framework satisfies $\varphi_2$ while remaining in the safe region (the green circle). }
\vspace{-5pt}
\label{fig:res2}
\vspace{-6pt}
\end{figure}

\end{Case Study}

\section{Conclusion and Future Work}
\label{sec:conclusion}

In this paper, we proposed a RNN framework to synthesize feedback control policies for a system under STL specifications. We used CBF to modify the control policies predicted by the RNN to guarantee safety, even in cases where safety constraints were unknown during the RNN training phase. We showed that our proposed RNN-CBF solution can be executed in real-time, while guaranteeing safety and achieving high success rate for STL satisfaction. Future research investigates utilizing the proposed RNN framework in model-free reinforcement learning approaches for control synthesis under STL specifications.

\bibliographystyle{IEEEtran}
\bibliography{references}

\end{document}